\documentstyle[12pt] {article} 
\newcommand{\ket}[1]{|#1\rangle}
\newcommand{\bra}[1]{\langle#1|}

\def\eg{{\it e.g.}}
\def\ie{{\it i.e.}}

\def\thalf{{\textstyle{\frac{1}{2}}}}

\def\tquar{{\textstyle{\frac{1}{4}}}}
\def\pj{\hspace{-.26cm}}
\def\fpj{\hspace{-.7cm}}

\begin{document}
\title{Collective Excitations \\and Ground State Correlations}
\author{E.~Heinz, H.~M\"uther\\Institut f\"ur
Theoretische Physik, Universit\"at T\"ubingen\\72076 T\"ubingen,
Germany\protect\\ \\H.A. Mavromatis\\
Physics Dept., King Fahd University of Petroleum and
Minerals\\Dhahran 31261, Saudi Arabia\protect\\}
\maketitle
\thispagestyle{empty}
\begin{abstract}
A generalized RPA formalism is presented which treats pp and ph correlations
on an equal footing. The effect of these correlations on the
single-particle Green function is discussed and it is demonstrated that a
self-consistent treatment of the single-particle Green function is
required to obtain stable solutions. A simple approximation scheme is
presented which incorporates for this self-consistency requirement
and conserves the number of particles. Results of numerical calculations
are given for $^{16}$O using a G-matrix interaction derived from a realistic
One-Boson-Exchange potential.
\end{abstract}

\newpage
\section{Introduction}
A lot of effort has been made to evaluate the basic properties of
nuclei, like binding energies and radii of charge distributions, directly
from a realistic model of the nucleon-nucleon (NN) interaction which
describes the experimental NN scattering data. Already many years ago
it was realized that such investigations require a very careful
consideration of the short-range correlations, which are induced by the
strong short-range and tensor components of a realistic NN interaction.
Various techniques of many-body theory have been developed to account
for these particular correlations. Here we would like to mention the
Brueckner-Bethe hole line expansion \cite{day}, the variational
approach, in particular the extension to the method of correlated
basis functions \cite{pand,fanto}, the so-called ``exponential S''
method \cite{zabo} and the self-consistent Green function approach
\cite{martin,wim}.

Such microscopic calculations, however, had rather limited success. As
an example we refer to the Brueckner hole-line expansion applied to
nuclear matter, for which very detailed studies have been made
considering various models for a realistic NN interaction. It turned
out that Brueckner-Hartree-Fock (BHF) calculations employing a nuclear
interaction
with a weak tensor force were able to reproduce the empirical value for
the binding energy of nuclear matter but predicted a saturation density
which was too large by almost a factor 2. Other interactions with
stronger tensor forces lead to satisfying results for the saturation
density but underestimate the binding energy by around 5 MeV per
nucleon. This result, that microscopic many-body calculations employing
realistic NN interactions fail to reproduce binding energy and
saturation density, is commonly referred to as the problem of the
``Coester band'' in nuclear matter \cite{coester}. Similar results are
obtained also for the ground-state properties of finite nuclei
\cite{zabo,prog1,carlo}.

Various attempts have been made to cure this problem and some of them even
succeeded to determine a saturation point of nuclear matter which is
very close to the empirical point. One possibility is to introduce
many-body forces \cite{pand2}. Kuo and coworkers demonstrated that the
inclusion of particle-particle hole-hole (pphh) ring diagrams tends to
shift the saturation point off the Coester band towards the empirical
value \cite{kuo1,kuo2}. Another possible solution is the inclusion of
relativistic effects within the Dirac-BHF approximation. Motivated by
the success of the Walecka model \cite{serot} various groups tried to
incorporate the modification of the Dirac spinor for the nucleon, due to
the interaction with the other nucleons, in BHF calculations
\cite{rupr,shakin,malf}. With the inclusion of the relativistic effects
one even obtains results for the saturation properties of nuclear
matter which are in agreement with the empirical data \cite{BM84}.

The success of the Dirac-BHF approach cannot be extended to the
description of finite nuclei. The relativistic features tend to significantly
improve the agreement between calculated values for the energies and radii
and the corresponding experimental data \cite{fritz,boers}.
The remaining discrepancies (around 0.7 MeV per nucleon in energy and
around 0.2 fm for the radius of light nuclei like $^{16}O$ and
$^{40}Ca$), however,  are still too large to be accounted for.

This situation encourages the investigation of correlations beyond
those included in the Dirac-BHF approximation. Such correlations should
have a sizeable effect on the groundstate properties of finite nuclei
but should not spoil the success of Dirac-BHF in nuclear matter. One
can expect that differences of this kind show up in correlations, which are
due to surface effects in finite nuclei. Such correlations are
typically described in terms of particle-hole (ph) excitations.
Therefore our studies should incorporate the correlation effects
contained in the particle-hole random phase approximation (ph RPA).
Furthermore we would also like to
account for the effects of pphh ring diagrams. As discussed above,
these correlations show very desirable effects in studies of nuclear
matter. Therefore it is interesting to investigate these correlations
for finite nuclei as well.

A ``super-RPA'' (SRPA) technique which allows the study of ph, pphh and
the possible
interference between these kinds of correlations was presented
a few years ago \cite{ellis1}. It has been possible to solve the
non-linear equations resulting from this SRPA approach for a small
model space or by an artificial weakening of the residual interaction. No
solution could be obtained for larger model-spaces using realistic
interactions. In the present investigation we would like to demonstrate
that this failure of the SRPA can be cured by a consistent
definition of the single-particle propagator.

After this introduction we summarize in section 2 of this paper the basic
ingredients of the SRPA and recall the equations to calculate
expectation values for the correlation energy and the occupation
probabilities. The self-consistent definition of the single-particle
propagator is discussed in section 3. As an example for an application
of the SRPA we present in section 4 results on $^{16}O$ employing
various realistic One-Boson-Exchange (OBE) interactions in different
model-spaces. The main conclusions are summarized in the final section.

\section{Ground-State Correlations in SRPA}
In this section we will briefly review the SRPA formalism, treating
particle-hole (ph), particle-particle (pp) and hole-hole (hh)
on the same level.
We shall work in a basis which is not angular-momentum
coupled, since this is more transparent. More details and also the
equations for states coupled to angular momenta are given in
ref.\cite{ellis1}. As usual the Hamiltonian
is broken into
an unperturbed part and a perturbation, $H=H_0+V$, where $H_0$ has
eigensolutions
\begin{equation}
H_0\ket A=\epsilon_A\ket A\;.
\end{equation}
We restrict our attention to a closed shell system for which we can use
standard non-degenerate perturbation theory,
see \eg\ Brandow \cite{baird}. Denoting the unperturbed closed shell
wavefunction by $\ket{\phi_0}$ and noting that we are interested in
2p-2h correlations, the true wavefunction can be written in the form
\begin{equation}
\ket{\psi}=\exp\Bigl[\sum_{ab\alpha\beta}t_{ab\alpha\beta}\,a^{\dagger}_a
a^{\dagger}_b a^{\phantom{\dagger}}_{\beta}
a^{\phantom{\dagger}}_{\alpha}\Bigr]\ket{\phi_0}\;,
\label{th1}
\end{equation}
where we use latin (greek) letters to designate states which are
unoccupied (occupied) in $\ket{\phi_0}$. A single letter here is used to
represent all the quantum numbers needed to specify an orbital.

We are then required to set up equations for the coefficients $t$.
The notation for the basic interactions $V$ that we have at hand are
standard: those that create (or destroy) a 2p-2h
state we denote by $B$, those that scatter a pair
of particles (or holes) we label ${\cal A}$ and those that scatter a p-h
pair we label $A$.
We can set up a multiple scattering series for $t$ in the form
\begin{eqnarray}
t_{ab\alpha\beta}&\pj=&\pj\frac{1}{\epsilon_{ab\alpha\beta}}
\sum_{{\rm repeated}\atop{\rm indices}}\biggl\{\tquar
B_{ab\alpha\beta}+\thalf{\cal A}_{abcd}t_{cd\alpha\beta}
+\thalf t_{ab\gamma\delta}{\cal A}_{\gamma\delta\alpha\beta}\nonumber\\
&&\hspace{2.1cm}-A_{a\gamma c\alpha}\left(t_{cb\gamma\beta}-t_{cb\beta\gamma}
-t_{bc\gamma\beta}+t_{bc\beta\gamma}\right)\nonumber\\
+\thalf(&&\fpj t_{ac\alpha\gamma}-t_{ac\gamma\alpha}-t_{ca\alpha\gamma}
+t_{ca\gamma\alpha})B_{\gamma\delta cd}\left(t_{bd\beta\delta}
-t_{bd\delta\beta}-t_{db\beta\delta}+t_{db\delta\beta}\right)\nonumber\\
&&\hspace{6cm}+t_{ab\gamma\delta}B_{\gamma\delta ef}t_{ef\alpha\beta}\biggr\}\,,
\label{srpa1}
\end{eqnarray}
where the energy
denominator $\epsilon_{ab\alpha\beta}=\epsilon_\alpha+\epsilon_\beta
-\epsilon_a-\epsilon_b$. Consider iterating this equation. The lowest order
contribution is simply the first term in the braces which
creates the 2p-2h state. Subsequent iterations build in the processes
shown in Fig. 1 (a) --- assuming $t$ is given in order $n$, order $(n+1)$ is
obtained by allowing an additional pp interaction (term 2 in braces), or hh
interaction (term 3), or one of the four possible ph interactions (term 4).
In addition we must allow for ``backward-going" $B$-vertices, as shown in
Fig. 1(b), which may be of ph type (term 5) or pp-hh type (term 6).
Here we have used the factorization theorem \cite{baird} to render the
energy denominators of the left and right boxes independent.
Thus the
wavefunction diagrams generated are of the general type indicated in
Fig. 1 (c) and, as advertised, we include diagrams with all orientations
of arrows on the lines subject to the requirement that $V$ does not
create particles, \ie, two arrows point towards and two away from each vertex.

Because of the antisymmetry of fermions, the only linear combination of
coefficients $t$ that enters is
\begin{equation}
t_{ab\alpha\beta}-t_{ab\beta\alpha}-t_{ba\alpha\beta}+t_{ba\beta\alpha}
\equiv2K_{ab\alpha\beta}\;.
\end{equation}
It is useful to exploit this in actual calculations in order to
cut down storage
requirements and it is simple to recast Eqs. (\ref{th1}) 
and (\ref{srpa1}) in terms of $K$:
\begin{equation}
\ket{\psi}=\exp\Bigl[\thalf\sum_{ab\alpha\beta}K_{ab\alpha\beta}\,a^{\dagger}_a
a^{\dagger}_b a^{\phantom{\dagger}}_{\beta}
a^{\phantom{\dagger}}_{\alpha}\Bigr]\ket{\phi_0}\;,
\label{th2}
\end{equation}
\begin{eqnarray}
K_{ab\alpha\beta}&\pj=&\pj\frac{1}{\epsilon_{ab\alpha\beta}}
\sum_{{\rm repeated}\atop{\rm indices}}\biggl\{\thalf
B_{ab\alpha\beta}+\thalf{\cal A}_{abcd}K_{cd\alpha\beta}
+\thalf K_{ab\gamma\delta}{\cal A}_{\gamma\delta\alpha\beta}\nonumber\\
&&\hspace{.58cm}-A_{a\gamma c\alpha}K_{cb\gamma\beta}+
A_{a\gamma c\beta}K_{cb\gamma\alpha}+A_{b\gamma c\alpha}K_{ca\gamma\beta}
-A_{b\gamma c\beta}K_{ca\gamma\alpha}\nonumber\\
&&\hspace{1.58cm}+2K_{ac\alpha\gamma}B_{\gamma\delta cd}K_{bd\beta\delta}
-2K_{ac\beta\gamma} B_{\gamma\delta cd}K_{bd\alpha\delta}\nonumber\\
&&\hspace{5.2cm}+\thalf K_{ab\gamma\delta}B_{\gamma\delta ef}K_{ef\alpha\beta}
\biggr\}\;.\label{srpa}
\end{eqnarray}

The ground state correlation energy is obtained by closing off the
wavefunction diagram of Fig. 1 (c) via a $B$-vertex.
Thus
\begin{eqnarray}
\Delta E_{\rm corr}&\pj=&\pj\bra\phi V\ket\psi\nonumber\\
&\pj=&\pj\sum_{ab\alpha\beta}B_{\alpha\beta ab}t_{ab\alpha\beta}
\equiv\thalf\sum_{ab\alpha\beta}B_{\alpha\beta ab}K_{ab\alpha\beta}\;.
\end{eqnarray}

For the norm of the correlated ground state wavefunction and the single particle
occupation probabilities, we generalize the approach discussed in Ref.
\cite{ellis2}. The norm of $\ket{\psi}$ can be written
\begin{equation}
\langle\psi|\psi\rangle=\exp\left\{\sum_{ab\alpha\beta \atop N}
\frac{L_{ab\alpha\beta}(N)}{N}K_{ab\alpha\beta}\right\}\;,
\label{n1}
\end{equation}
where the quantity in braces gives the sum of all linked normalization diagrams
\cite{baird} and the exponential form follows from considering all possible
products of linked diagrams. For $N=1$ we have
\begin{equation}
L_{ab\alpha\beta}(1)=K_{ab\alpha\beta}\;.
\end{equation}
This clearly arises from expanding the exponential form given for $\ket{\psi}$
in Eq. (\ref{th2}) and taking the contribution 
to $\langle\psi|\psi\rangle$ which
contains
2 $K$-coefficients. To include the linked contribution from 4,6, \ldots\
$K$-coefficients, corresponding to $N=2,3,\ \ldots$ , we define an iterative
equation for $L$:
\begin{eqnarray}
&&\fpj L_{ab\alpha\beta}(N)=\sum_{{\rm repeated}\atop{\rm indices}}\biggl\{
L_{ab\gamma\delta}(N-1)K_{cd\gamma\delta}K_{cd\alpha\beta}\nonumber\\
&&\hspace{.45cm}+2L_{ad\alpha\delta}(N-1)K_{cd\gamma\delta}K_{bc\beta\gamma}
-2L_{bd\alpha\delta}(N-1)K_{cd\gamma\delta}K_{ac\beta\gamma}\nonumber\\
&&\hspace{.45cm}-2L_{ad\beta\delta}(N-1)K_{cd\gamma\delta}K_{bc\alpha\gamma}
+2L_{bd\beta\delta}(N-1)K_{cd\gamma\delta}K_{ac\alpha\gamma}\biggr\}.
\label{l1}
\end{eqnarray}
Here $L(N)$ is built by joining $L(N-1)$ to one $K$ coming from the bra and
another coming from the ket. Thus $L(N)$ corresponds to a chain of $(2N-1)$
coefficients $K$ which is linked
together in a manner consistent with our previous diagrams. The chain is finally
closed off by taking $LK$ in the braces of Eq. (\ref{n1}). 
The first term on the right
in Eq. (\ref{l1}) connects via pp and hh pairs; each $K$ carries a factor of
$\thalf$, but each pair ($cd$ or $\gamma\delta$) can be contracted in two
ways, hence the net factor is unity. The remaining four terms connect via
ph pairs; the 4 ways of picking $ab\alpha\beta$ are shown explicitly. We may
also interchange the particle and/or hole labels on the $K$'s which, by
symmetry, gives a factor of 16; against this must be set the factor of
$\thalf$ carried by each $K$ and a further factor of $\thalf$ because our
interchanges implicitly include $KKL$ as well as $LKK$ and these are
equivalent.
Thus there is a net factor of 2. We should also comment on the factor of
$N^{-1}$ in Eq. (\ref{n1}). 
In joining together the $K$'s, there are $N!$ permutations
in the bra, but $(N-1)!$ permutations in the ket to yield distinct (and
identical) results. Combining with the factor of $(N!)^{-2}$ arising from the
expansion of the exponentials, we obtain a net factor of $N^{-1}$.

The single particle occupation probabilities can be obtained by evaluating
the expectation value of the operator
\begin{equation}
{\cal N}_P=a^{\dagger}_Pa^{\phantom{\dagger}}_P\;,
\end{equation}
where the orbital $P$ is unoccupied in $\ket{\phi_0}$. This operator will
pick out the particular orbital $P$ from the chain of $K$-coefficients
discussed above and since there are $2N$ possibilities, each of which gives
the same result, we obtain
\begin{equation}
         n_{P}             \equiv\frac{\langle\psi|{\cal N}_P|\psi\rangle}
{\langle\psi|\psi\rangle}=2\sum_{b\alpha\beta\atop N}
L_{Pb\alpha\beta}(N)K_{Pb\alpha\beta}\;.\label{occu}
\end{equation}
A completely analogous expression can be given for the hole occupation
probabilities.

More details on the SRPA equations, the embedding of various
approximations and the coupling to states of good angular momentum are
given in ref.\cite{ellis1}.

\section{Correlations and Single-Particle Properties}

The correlations considered in the SRPA formalism presented in the
preceding section are reflected in occupation probabilities which
deviate from the occupation numbers one and zero of the independent
particle model. This implies that the Lehmann representation of the
single-particle Green function is not any longer given by only one pole
but must be presented in the general form (see e.g.\cite{wim})
\begin{equation}
g(\alpha,\omega) = \int_{-\infty}^{\epsilon_{F}} d\omega '\,
\frac{S_{h}(\alpha,\omega ')}{\omega - \omega ' + i\eta} \; + \;
\int_{\epsilon_{F}}^{+\infty} \frac{S_{p}(\alpha,\omega ')}
{\omega - \omega ' - i\eta} \; .\label{lehmann}
\end{equation}
In this equation the index $\alpha$ refers to the symmetry quantum
numbers of the single-particle state under consideration and $\omega$
is the energy variable in the single-particle Green function
$g(\alpha,\omega)$. The hole spectral function $S_{h}(\alpha,\omega)$
is related to the occupation number of the single-particle state
$\alpha$ in the ground state of the nucleus by
\begin{eqnarray}
n_{\alpha} & = &
\frac{<\psi \vert a_{\alpha}^\dagger a_{\alpha}\vert \psi >} {<\psi
\vert \psi > }\nonumber \\
& = & \int_{-\infty}^{\epsilon_{F}} d\omega \,
S_{h}(\alpha,\omega )
\end{eqnarray}
where the integral is restricted to energies $\omega$ below the Fermi
energy $\epsilon_{F}$. Within the framework of the Green function
formalism the single-particle Green function is usually determined by
considering a certain approximation for the self-energy of the nucleons
with quantum numbers $\alpha$ and solving a Dyson equation using this
approximation for the self-energy. A problem of this Green function
approach is that only very specific approximations for the self-energy
lead to a number conserving approach, namely
\begin{equation}
\sum_{\alpha} (2j_{\alpha} +1) n_{\alpha} = N
\end{equation}
which means that the sum on all occupation probabilities multiplied by
the degeneracy of the state $\alpha$ (here $2j_{\alpha}+1$) yields the
required particle-number $N$. It turns out that rather sophisticated
self-consistency requirements have to be fulfilled for any
approximation to the self-energy beyond the Hartree-Fock approach
\cite{baym}.

The hole spectral functions $S_{h}(\alpha,\omega )$ and
particle spectral functions $S_{p}(\alpha,\omega )$ can be used to
define mean single-particle energies below
\begin{equation}
\epsilon_{<,\alpha} =\frac{1}{n_{\alpha}} \int_{-\infty}^{\epsilon_{F}}
d\omega \,\omega S_{h}(\alpha,\omega )
\end{equation}
and above the Fermi energy
\begin{equation}
\epsilon_{>,\alpha} =\frac{1}{1-n_{\alpha}} \int_{\epsilon_{F}}^{+\infty}
d\omega \,\omega S_{p}(\alpha,\omega )\; .
\end{equation}
Furthermore it can be shown that for a very general approximation in
the self-energy these mean values are related to the Hartree-Fock
single-particle energy $\epsilon_{\alpha}^{HF}$ by \cite{skoura1}
\begin{equation}
n_{\alpha}\epsilon_{<,\alpha} \, + \, (1-n_{\alpha})\epsilon_{>,\alpha}
\; = \;\epsilon_{\alpha}^{HF} \; .\label{epsi}
\end{equation}
Using the nomenclature introduced so far we can now define an
approximation for the single-particle Green function which is a first
step beyond the Hartree-Fock approximation by simplifying
eq.(\ref{lehmann}) to
\begin{equation}
g^{(1)}(\alpha,\omega ) = \frac{n_{\alpha}}{\omega-\epsilon_{<,\alpha}
+ i\eta} + \frac{1-n_{\alpha}}{\omega-\epsilon_{>,\alpha} - i\eta}
\label{green1}
\end{equation}
This means that a nucleon in a single-particle state $\alpha$ is
propagating as a hole with probability $n_{\alpha}$ at an energy
$\epsilon_{<,\alpha}$ and propagating as a particle with probability
$(1-n_{\alpha})$ at an energy $\epsilon_{>,\alpha}$.

In our investigation we want to consider the interplay between the
correlations as calculated in the SRPA approach discussed in the
previous section and the single-particle propagator using the
approximation of eq.(\ref{green1}). This means that we employ the
occupation numbers calculated in SRPA according to eq.(\ref{occu}).
These occupation numbers $n_{\alpha}$ and the relation (\ref{epsi})
are used to determine the energies $\epsilon_{<,\alpha}$ and
$\epsilon_{>,\alpha}$. This is done in the following way: For states
$\alpha$, which are mainly occupied ($n_{\alpha}>0.5$) we assume for
the energy of the particle component $\epsilon_{>,\alpha}$ (which is of
minor importance for this orbit), the value
determined by the BAGEL technique of ref.\cite{skoura2} for the
corresponding nucleus and hamiltonian. The energy of the dominant hole
component $\epsilon_{<,\alpha}$ is then determined according to 
(\ref{epsi}). For orbits which are predominantly particle states
($n_{\alpha}<0.5$) we proceed in the analogous way and determine the
energy for the propagation of the hole state from ref.\cite{skoura2},
while the energy of the particle part ($\epsilon_{>,\alpha}$) is
derived from (\ref{epsi}).

The modified single-particle Green function of eq.(\ref{green1}) is
then used in the SRPA equations. This means that all summations in
e.g. eq.(\ref{srpa}) which were restricted to particle states now have
to consider all orbits $\alpha$ with a weight $(1-n_{\alpha})$ and an
energy of $\epsilon_{>,\alpha}$. The corresponding statement holds for
the hole orbits. This solution of the modified SRPA equations leads to
new occupation probabilities, which give rise to different
single-particle energies. This procedure is repeated until a
self-consistent solution is obtained.

It is evident that this self-consistent determination of the
single-particle Green function yields a stabilizing effect for the
solution of the SRPA equations: If correlation effects are large, the
occupation probabilities will deviate strongly from the values of
the independent particle model. According to eq.(\ref{epsi}) this gives
rise to hole-energies $\epsilon_{<,\alpha}$ which are significantly below
and particle-energies $\epsilon_{>,\alpha}$ significantly above the
corresponding Hartree-Fock value $\epsilon_{\alpha}^{HF}$. Therefore
one obtains an additional gap between particle- and hole-energies. This
change in the single-particle spectrum tends to reduce the correlation
effects, calculated in SRPA. This is what we call the stabilizing
effect of the single-particle Green function.

\section{Results and Discussion}
\subsection{SRPA and Independent Particle Model}
As a first step in our discussion of results we would like to recall
some of the features and the problems of the SRPA approach
already discussed in ref.\cite{ellis1}. In this first part we consider
the SRPA introduced in section 2, i.e.~assuming single-particle
propagators of the kind of the independent particle model (IPM). This
means that the different single-particle orbits of our model space
are either hole-states and completely occupied or particle-states
(completely unoccupied).

As in \cite{ellis1} we will consider the nucleus $^{16}$O assuming for
the first part of this discussion a
very small model space. The summation on hole-states shall be
restricted to the states of the p-shell while the orbits of the 1s0d
shell are considered for the particle states. The matrix elements of
the NN interaction are evaluated in a basis of oscillator states
(oscillator parameter $b$=1.76 fm) by solving the Bethe-Goldstone
equation \cite{sauer} for the OBE potentials $``A''$, $``B''$ and
$``C''$ defined in table A.1 of \cite{rupr}. More details on the
evaluation of these matrix elements and a modification of the
interaction which ensures that the chosen oscillator basis is identical
to the self-consistent Hartree-Fock (HF) basis are discussed in
\cite{skoura1}.

One can determine the HF single-particle energies for these three
sets of G-matrix elements and try to solve the non-linear SRPA equations
by an iterative procedure. It turns out, however, that such an
iterative scheme does not converge. The SRPA equations can be
simplified to account only for correlations obtained already in the
conventional particle-hole (ph) RPA or the particle-particle hole-hole
(pphh) RPA approach (see detailed discussion in \cite{ellis1}).
The iterative scheme for the solution of these
reduced SRPA equations converges and yields the same result for the
correlation energy as can be obtained by other techniques
\cite{brown,yang,harry1}.

In the case of a model space, which contains an inert core (like the
$^{4}$He core in the model space presently considered, the HF choice
for the single-particle energies may not be appropriate. Therefore in
\cite{ellis1} we have employed a set of single-particle energies, which
is not really self-consistent. Using the same set of single-particle
energies we obtain results for the correlation energies using various
approximations and interactions as displayed in table \ref{tab:tab1}.
These results
are very similar to results already discussed in \cite{ellis1}, the
differences to the results reported before are due to a different
choice of the Pauli operator and starting energy in the solution of the
Bethe-Goldstone equation.

The total correlation energy is split into the contribution due to the
term of second order in the residual interaction, which is part of the
correlation energy in all the approaches under consideration and the
correlation energy due to the terms of third and higher order in the
residual interaction. In table \ref{tab:tab1} we compare correlation energies
resulting from the ph-RPA (with inclusion of exchange terms, see
\cite{ellis1}), the pp-hh RPA and the SRPA. As it has already been
discussed in \cite{ellis1}, one can draw the following conclusions:
\begin{itemize}
\item The correlation energy
obtained in SRPA is larger than the sum of correlation energies
resulting from ph-RPA and pp-hh RPA. This additional energy is due to
terms which contain a residual interaction between ph excitations but
also between 2 particle or 2 hole states.
\item The correlation energy due to
the terms of third and higher order are essentially as large as the
correlation energy of the second-order term.
\item The residual interaction is decreasing from OBE potential $A$ to
$B$ to $C$. This can be deduced from the terms of second order but also
from the total correlation energies. This is in line with the fact that
the potential $A$ also yields the largest binding energy in BHF
calculations of nuclear matter \cite{rupr} or finite nuclei
\cite{carlo}.
\end{itemize}
At first sight all these results seem to be reasonable. A more detailed
discussion and investigation, however, gives rise to some doubts:
\begin{description}
\item[(i)] It has already been mentioned that problems arise if HF 
single-particle energies are used in the SRPA approach.
\item[(ii)] A closer inspection of the wavefunction shows that the
correlation determined by SRPA are very strong, leading to occupation
probabilities for the hole states which are small already for the
OBE potential $C$ (0.814 and 0.822 for $p_{3/2}$ and $p_{1/2}$ shell,
respectively) and extremely small for OBE $A$ (0.276 and 0.375).
\item[(iii)] Enlarging the model space to include all hole states of
$^{16}$O and the 1s0d plus 1p0f shells for the particle states leads to
SRPA equations, which cannot be solved neither by employing the
single-particle energies of \cite{ellis1} nor using the HF
single-particle energies.
\end{description}
It is worth noting that the iterative solution of the SRPA equations
converges also in the large model-space if we assume HF single-particle
energies and reduce the SRPA equations to the ph-RPA or pp-hh RPA
approach. The correlation energy is the same as the one obtained from
a diagonalisation of the corresponding RPA equation. If, however, we
keep the single-particle energies fixed and continuously enhance the
residual interaction by multiplying the matrix elements with a constant
($\lambda >$ 1), we observe that at a critical value of $\lambda$ the
iteration of the non-linear equation does not converge any more. At the
same value of $\lambda$ the diagonalisation of the corresponding RPA
equations yields complex eigenvalues.

Therefore we conclude that the HF solution is stable with respect to
ph-RPA and pp-hh RPA correlations. This is reflected in our
calculations by the fact that RPA calculations using self-consistent
single-particle energies yield real energies for the RPA phonons and
iterative solutions for the corresponding non-linear equations and, of
course, a feature of the HF approximation \cite{thoul}. The fact that
the complete SRPA equations do not show an iterative solution can be
interpreted to indicate that the HF state is not stable with respect to
the excitations of the SRPA kind. The SRPA calls for a better
approximation of the single-particle Green function than the one derived
from HF.

A very pragmatic way to obtain solution of the full SRPA equations is
to introduce a gap between the energies of particle and hole-states.
Indeed, shifting e.g.~all energies for particle states by a constant of
around 3 MeV leads to stable solutions even for the large model space.
This shift is rather arbitrary. Therefore we prefer to investigate the
modification of the single-particle Green function introduced in
section 3, which establishes a self-consistent link between the
evaluation of the correlations and the definition of the
single-particle properties.

\subsection{``Self-Consistent'' SRPA}
As we expected from our discussion in section 3, the
``self-consistent'' choice of the single-particle propagator according
eq.(\ref{green1}) yields a stabilization of the SRPA equations. With
this choice we obtain solutions of the non-linear SRPA equations for
the small and large model-spaces for all interactions considered.
Results for the correlation energies are displayed in table
\ref{tab:tab2} considering the same approximations, interactions and
model-spaces as in table \ref{tab:tab1}. 

The occupation numbers for the single-particle orbits $n_{\alpha}$
depend on the approximation considered. In table \ref{tab:tab3}
occupation numbers are listed, which were calculated for the OBE $B$
potential using different approximations and model-spaces. Considerable
deviations from the occupations of the IPM are observed. These
deviations are slightly larger if one uses the interaction OBE $A$ and
slightly smaller for the interaction $C$. As the occupation numbers
depend on the method considered, also the single-particle propagators
defined in eq.(\ref{green1}) will be different. Consequently for these
self-consistent calculations also the contribution of the second order
term will be different for the different approaches. In table \ref{tab:tab2}
we give for each approximation the contribution of this second order
term calculated with the self-consistent single-particle propagator and
separately the sum of all higher order terms.

The stabilization of the correlation effects due to the self-consistent
choice of the single-particle propagator is well documented by the
contribution to the energy of second order in the residual interaction. For
approximations which yield large correlation effects, like the SRPA,
one obtains sizeable deviations of the occupation probabilities from
the IPM (see table \ref{tab:tab3}) and contributions to the binding
energy from the second order term, which are considerably smaller than
the second order terms calculated without readjustment of the
single-particle features (see table 1). For approaches like RPA, which
yield weaker correlation effects, the quenching of the second order
term due to the single-particle propagator is weaker. Therefore the
more correlations are taken into account (SRPA as compared to RPA) the
smaller the contribution of the second order term. Also the
contributions of the terms of third and higher order in the interaction
are reduced due to the self-consistent single-particle propagator. The
propagator effect, however, is generally weaker than for the second
order terms.

It is interesting to note that the sum of second order plus all higher
order terms yields a correction to the binding energy, which is almost
independent on the approximation considered. For the small model-space
we obtain around 9 MeV, while the calculations in the large model-space
yield around 19 MeV. This does not imply that
correlations beyond RPA are negligible if the single-particle propagator
is chosen in a self-consistent manner. Indeed the contribution to the
binding energy originating from third and higher order terms is
increasing going from RPA via pp-hh RPA to SRPA. This enhancement of
the higher order terms is partly canceled by reduction of the second
order terms as already discussed above. The importance of correlations
beyond RPA is also reflected by the occupation probabilities displayed
in table \ref{tab:tab3}. The deviations from the Hartree-Fock values
observed in the SRPA are much larger than those obtained in RPA (see
also Fig. 2). This
is true in particular for the 2 shells close to the Fermi level (the
$0p$ and the $1s0d$ shells). In this case the deviations obtained in
SRPA are larger than the sum of the deviations produced by RPA and
pp-hh RPA correlations.

The inspection of the correlation energies for the various model-spaces
(9 MeV for the small and 19 MeV for the large model-space) may cast
doubt on the convergence of the calculated correlation energy with
respect to the size of the model-space. It should be noted, however,
that the occupation probabilities obtained for the shells away from the
Fermi level are not very sensitive to the approximation considered. The
RPA and also the pp-hh RPA yields results for the $0s$ and $1p0f$
shells, which are very similar as those obtained in SRPA. This
indicates that the interplay between the various correlations contained
in SRPA is of dominant importance only for the shells close to the
Fermi energy. Therefore the pp-hh RPA, which is the summation of all
particle-particle and hole-hole ladders, should provide a good
approximation for the correlations of these high energy particle-hole
excitations.

One possible way to account for these pp-hh correlations would be the
use of the self-consistent Green function method \cite{wim}. The Green
function approach, if considered beyond the Hartree-Fock approximation,
requires a very sophisticated self-consistency between the
approximation used for the 2-particle Green function and the
single-particle Green function to guarantee number conservation. The
approximation for the single-particle propagator, which we are using
here, is similar to the Green function approach (see discussion in
3.2). The present scheme, however, has the advantage that it always
yields the correct particle number.

Employing the self-consistent single-particle Green functions defined
in eq.(\ref{green1}) one can calculate the expectation value for
single-particle operators like the radius of the nucleon distribution.
The radii obtained for the various approximations, using the OBE
potentials $A$, $B$ and $C$ in
the large model-space are listed in table \ref{tab:tab4} and
compared to the result obtained in the Hartree-Fock approximation. 
As we discussed already in the beginning of this section, the
interactions have been modified to guarantee a Hartree-Fock solution with
a radius close to the experimental value \cite{skoura1}.
Therefore we should only discuss the modifications in the calculated radius
due to the correlation effects. One finds a small but non-negligible
enhancement for the calculated radius. As to be expected also for this
observable the effect of correlations is larger for the SRPA than for
the 2 RPA approximations. Also for the calculation of the radii, the
effects of correlations is slightly increasing going from potential
$A$ to $C$.

It is remarkable that the inclusion of SRPA
correlation increases the binding energy but also the value for the
calculated radius. This implies that calculated ground-state properties
are moved off the ``Coester-band'' towards the experimental point. This
effect is not very large but together with the features of the
Dirac-BHF approach it may be sufficient to yield results for the
ground-state of finite nuclei, which are in good agreement with the
empirical data.

\section{Conclusions}
A method is discussed which allows the non-perturbative evaluation of
correlation effects beyond the conventional particle-hole (ph) and
particle-particle hole-hole (pp-hh) Random-Phase approximations. It
turns
out that the resulting non-linear equations of this SRPA yield solutions
only, if the single-particle propagator or single-particle Green
functions are evaluated in a self-consistent way beyond the
Hartree-Fock approximation. A self-consistent scheme is developed,
which conserves the particle number, i.e.~the solutions for the
single-particle Green function yield the required particle number.

It turns out that the interference between ph and pp-hh correlations, as
described by SRPA, is rather important for correlation effects within
the major shells close to the Fermi-level. The correlation effects due
to configurations involving shells of deep-lying hole states or
high-energy particle states seem to be described well within the RPA.
The correlations increase the value for the calculated binding energy
and lead to an enhancement for the calculated radius. Therefore
correlations of the SRPA type combined with the features of the
Dirac-BHF approximation \cite{fritz} may produce results for the
ground-state properties of finite nuclei, which are in good agreement
with the experimental data.

This investigation has been done within the ``Graduiertenkolleg
Struktur und Wechselwirkung von Hadronen und Kernen''. The financial
support by the Deutsche Forschungsgemeinschaft (DFG Mu 705/5) and the
support of the ``King Fahd University of Petroleum and Minerals'' is
acknowledged. We are grateful to Professor Paul Ellis for useful
discussions

\clearpage
\begin{table}[h]
\caption{Correlation energies for $^{16}O$ obtained with fixed single-particle
energies as defined in ref.[20]. The correlation energy for a specific
approach is the sum of the term of second-order in the residual interaction
(first line labeled 2.order) plus the contribution of all higher order
terms which are listed in the lines identified by the various
approaches. Results are listed for the OBE potentials $A$, $B$ and $C$
defined in table A.1 of ref.[14] considering a small (0p and 1s0d shells)
and a large model space (all shells up to 1p0f shell). All energies are
given in MeV.}
\label{tab:tab1}
\begin{center}
\begin{tabular}{||c|rrr|rrr||}
\hline\hline
&&&&&&\\
& \multicolumn{3}{|c|}{small space}
&\multicolumn{3}{c||}{large space}\\
&\multicolumn{1}{|c}{OBE $A$}
&\multicolumn{1}{c}{OBE $B$}
&\multicolumn{1}{c|}{OBE $C$}
&\multicolumn{1}{c}{OBE $A$}
&\multicolumn{1}{c}{OBE $B$}
&\multicolumn{1}{c||}{OBE $C$}\\
&&&&&&\\
\hline
&&&&&&\\
2.order & -9.68 & -9.23 & -8.95 & -21.45 & -20.77 & -20.41 \\
ph RPA & -1.79 & -1.46 & -1.27 & -3.53 & -2.44 & -1.97 \\
pp-hh RPA & -3.36 & -2.93 & -2.71 & -8.65 & -7.56 & -7.06 \\
SRPA & -8.83 & -6.98 & -6.05 & div & div & div\\
&&&&&&\\
\hline\hline
\end{tabular}
\end{center}
\end{table}
\begin{table}[h]
\caption{Correlation energies for $^{16}O$ obtained with single-particle
propagators as defined in eq.(3.19). For each approximation the first line
gives the contribution of the terms of second order, whereas the second
line shows the sum of all terms of higher order in the residual
interaction. Further details see table 1}
\label{tab:tab2}
\begin{center}
\begin{tabular}{||c|rrr|rrr||}
\hline\hline
&&&&&&\\
& \multicolumn{3}{|c|}{small space}
&\multicolumn{3}{c||}{large space}\\
&\multicolumn{1}{|c}{OBE $A$}
&\multicolumn{1}{c}{OBE $B$}
&\multicolumn{1}{c|}{OBE $C$}
&\multicolumn{1}{c}{OBE $A$}
&\multicolumn{1}{c}{OBE $B$}
&\multicolumn{1}{c||}{OBE $C$}\\
&&&&&&\\
\hline
&&&&&&\\
2.order & -6.75 & -6.67 & -6.61 & -14.60 & -14.67 & -14.73 \\
ph RPA & -1.68 & -1.46 & -1.33 & -3.41 & -2.92 & -2.68 \\
&&&&&&\\
2.order & -6.34 & -6.27 & -6.21 & -12.76 & -12.85 & -12.88 \\
pp-hh RPA & -2.96 & -2.67 & -2.53 & -7.79 & -7.22 & -7.00 \\
&&&&&&\\
2.order & -4.77 & -4.85 & -4.90 & -8.87 & -9.51 & -9.85 \\
SRPA & -5.15 & -4.68 & -4.41 & -11.14 & -10.20 & -9.73 \\
&&&&&&\\
\hline\hline
\end{tabular}
\end{center}
\end{table}
\begin{table}[h]
\caption{Occupation probabilities for the single-particle orbits listed
in column 1, calculated from the particle numbers
of eq.(2.12). Results are presented for the NN interaction``B'', for
various approximations and model spaces (see caption of table 1). The
last column lists occupation probabilities, which are obtained if for
the SRPA approach in  the large model space the
sum on the right hand side of eq.(2.12) is restricted to $N=1$.}
\label{tab:tab3}
\begin{center}
\begin{tabular}{||c|r|rrr|r||}
\hline\hline
&&&&&\\
& \multicolumn{1}{|c|}{small space}
&\multicolumn{3}{c|}{large space}&\\
&\multicolumn{1}{c|}{SRPA}
&\multicolumn{1}{c}{ph RPA}
&\multicolumn{1}{c}{pp-hh RPA}
&\multicolumn{1}{c|}{SRPA}
&\multicolumn{1}{c||}{Approx} \\
&&&&&\\
\hline
&&&&&\\
$s_{1/2}$ &  - & 0.976 & 0.966 & 0.964 & 0.966 \\
$p_{3/2}$ & 0.918 & 0.942 & 0.935 & 0.842 & 0.894 \\
$p_{1/2}$ & 0.925 & 0.944 & 0.922 & 0.863 & 0.889 \\
&&&&&\\
$d_{5/2}$ & 0.045 & 0.027 & 0.028 & 0.078 & 0.051 \\
$1s_{1/2}$ & 0.019 & 0.017 & 0.027 & 0.045 & 0.036 \\
$d_{3/2}$ & 0.042 & 0.027 & 0.034 & 0.072 & 0.053 \\
$f_{7/2}$ & - & 0.002 & 0.003 & 0.003 & 0.003 \\
$1p_{3/2}$ & - & 0.008 & 0.010 & 0.011 & 0.011 \\
$f_{5/2}$ & - & 0.004 & 0.007 & 0.006 & 0.006 \\
$1p_{1/2}$ & - & 0.007 & 0.009 & 0.010 & 0.009 \\
&&&&&\\
\hline\hline
\end{tabular}
\end{center}
\end{table}

\begin{table}[h]
\caption{Results for the radius (in [fm]) of $^{16}$O calculated in various
approximations are presented using the three different OBE potentials
A, B and C.
The interactions have been adjusted to obtain the same result in the
Hartree-Fock (HF) approximation. Correlation effects have been
calculated in the large model-space.}
\label{tab:tab4}
\begin{center}
\begin{tabular}{||c|rrr||}
\hline\hline
&&&\\
& \multicolumn{1}{|c}{A}
&\multicolumn{1}{c}{B}
&\multicolumn{1}{c||}{C}\\
&&&\\
\hline
&&&\\
HF & 2.634  & 2.634 & 2.634 \\
ph RPA & 2.673 & 2.671 & 2.671 \\
pp-hh RPA & 2.684 & 2.683 & 2.682 \\
SRPA & 2.732 & 2.718 & 2.712 \\
&&&\\
\hline\hline
\end{tabular}
\end{center}
\end{table}
\clearpage
{\bf Figure Caption}
\begin{itemize}
\item[Figure 1:] Iteration of Eq.(\ref{srpa1}) including (a)
forward-going and (b) backward-going vertices. In (c) we show a general
diagram for the wavefunction.
\bigskip
\item[Figure 2:] The occupation probabilities in $^{16}$O for the orbits
in the $0s$ to $1s0d$ shell are represented in terms of columns. The
results obtained by various approximations are distinguihed by the
shading of the bars. All results were obtained for OBE potential $C$.
\end{itemize}
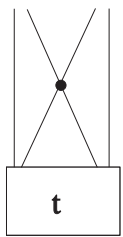
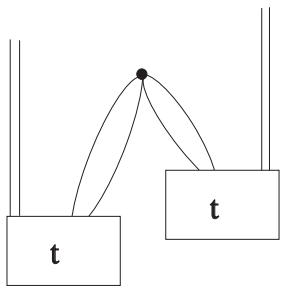
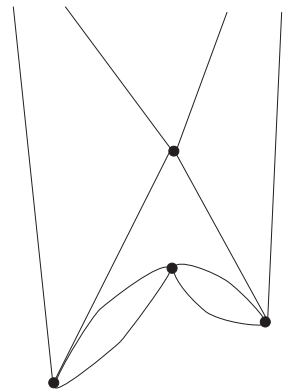
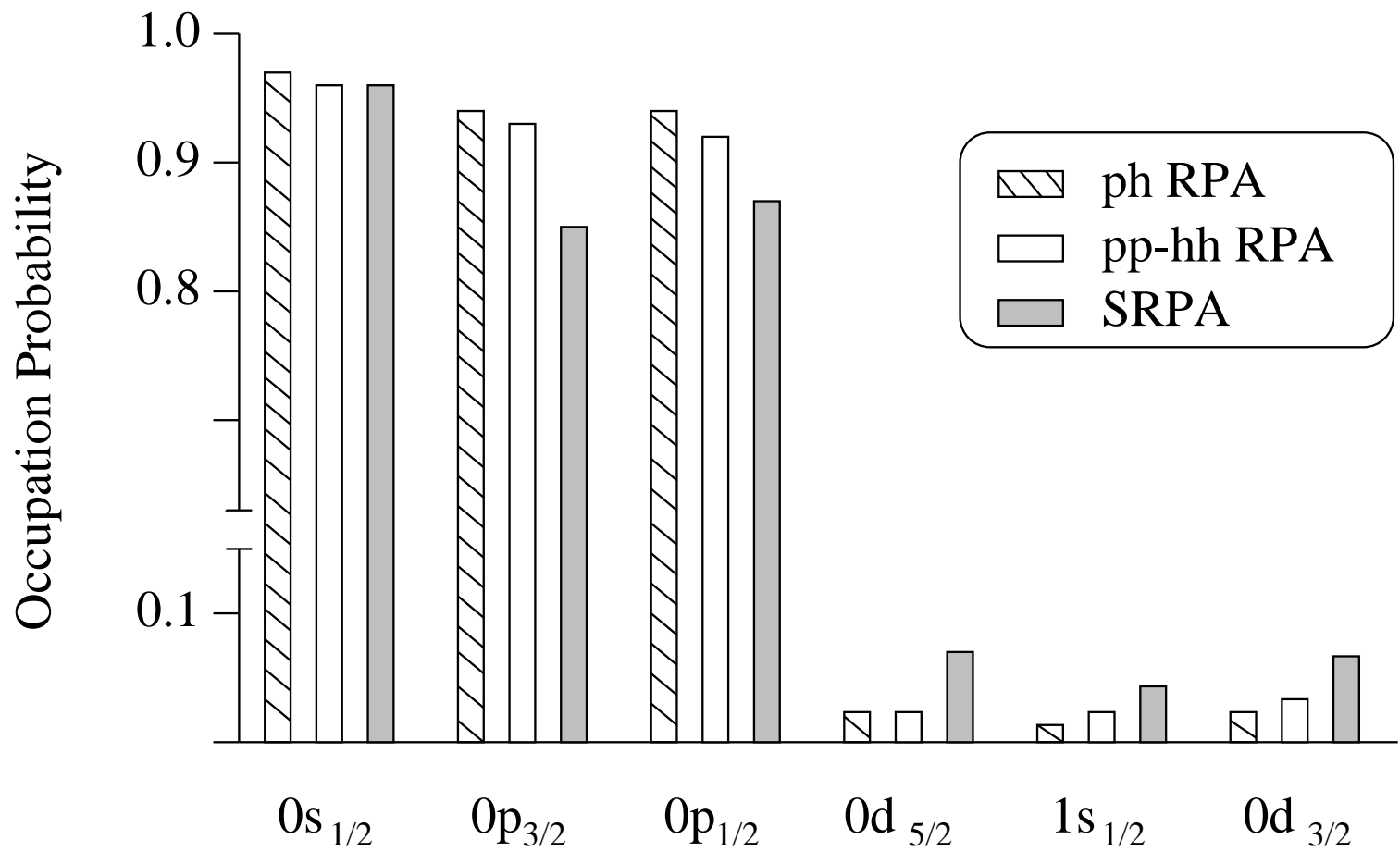
\end{document}